\documentclass[fleqn,10pt]{wlscirep}
\usepackage{amsmath}
\usepackage{dcolumn}
\usepackage{amssymb,amsmath}
\usepackage{color}
\usepackage{epsfig}
\usepackage{graphicx}
\usepackage{epstopdf}

\newcommand{\blue}{\textcolor{blue}}

\title{Efficient X-ray generation from gold colloidal solutions}

\author[1,*]{Koji Hatanaka}
\author[2]{Matteo Porta}
\author[1]{Frances Camille P. Masim}
\author[1]{Wei-Hung Hsu}
\author[2]{Mai Thanh Nguyen}
\author[2]{Tetsu Yonezawa}
\author[3,4]{Armandas Bal\v{c}ytis}
\author[3,5]{Saulius Juodkazis}

\affil[1]{Research Center for Applied Sciences, Academia Sinica,
Nangang, Taipei 11529, Taiwan} \affil[2]{Department of Materials
Science, Graduate School of Engineering, Hokkaido University,
Sapporo 060-8628, Japan} \affil[3]{Faculty of Science, Engineering
and Technology, Swinburne University of Technology, Hawthorn, VIC
3122, Australia} \affil[4]{Institute of Physics, Center for
Physical Sciences and Technology, Vilnius, LT-02300, Lithuania}
\affil[5]{Center for Nanotechnology, King Abdulaziz University,
Jeddah 21589, Saudi Arabia}

\affil[*]{kojihtnk@gate.sinica.edu.tw}

\begin{abstract}
Hard X-ray generation for Au nanoparticle dispersion was
systematically investigated for different particle diameters
ranging from 10 to 100~nm with a narrow size distribution of
$\pm$2\%. Scaling law of X-ray generation is close to a 6-photon
process before the onset of saturation for excitation by 45~fs
laser pulses with central wavelength of 800~nm. This is consistent
with bulk plasmon excitation at $\lambda_p^{bulk} \simeq 138$~nm.
The longitudinal E-field component due to nanoparticle focusing is
responsible for the excitation of the longitudinal bulk plasmon.
The proposed analysis also explains X-ray emission from water
breakdown via an electron solvation mechanism at $\sim$6.2~eV. The
most efficient emission of X-rays was observed for 40$\pm$ 10~nm
diameter nanoparticles which also had the strongest UV extinction.
X-ray emission was the most efficient when induced by pre-chirped
370$\pm$20~fs laser pulses and exhibited the highest intensity at
a negative chirp.
\end{abstract}
\begin{document}

\flushbottom \maketitle
%
%
\thispagestyle{empty}

\section*{Introduction}

Compact and efficient hard X-ray sources can revolutionize
biological and medical fields as well as practical applications.
Liquid targets for X-ray generation by optically induced breakdown
are attractive due to the possibility to tailor the X-ray spectrum
via choice of ion content of the micro-jet or droplet irradiated
by high intensity laser pulses at $\sim
10^{15}$~W/cm$^2$~\cite{NewHatanaka}. At even higher intensities
approaching $\sim 10^{20}$~W/cm$^2$ (in vacuum), particle
acceleration beams for electrons and protons up to MeV energies
are demonstrated using peta-watt lasers~\cite{Murakami}.

In the case of ultra-short laser pulses, a wide range of control
parameters over light-matter interaction exist, some of which, are
specific to the ultrashort laser pulses: spatial and temporal
chirp. Polarisation, optical nonlinearities of irradiated
materials, different contributions of E-field components due to
peculiarities of focusing at the near-field, provide ways to
enhance electron temperature and concentration for brighter X-ray
emission as briefly introduced below.

It has been demonstrated that the resonant absorption mechanism is
responsible for hard X-ray generation in the case of metal-doped
glasses~\cite{08jncs5485} and water jets~\cite{jets,08oe12650}
irradiated by $\sim 150$~fs pulses. By introduction of a negative
chirp and choosing p-polarized pulses, X-ray generation is
enhanced by almost an order of magnitude~\cite{08oe12650}. For 5-6
times shorter pulses, the multi-photon ionization (MPI) becomes
comparable to or more efficient than impact ionization (IMP), and
even higher electron densities and temperatures are
reached~\cite{Gam,Rode,10oe10209}. In the case of X-ray generation
from wide-bandgap undoped dielectrics such as pure silica,
sapphire, or water with high dielectric strength against
dielectric breakdown, the polarization dependent rates of MPI
become important, especially for short sub-50~fs pulses. Surface
breakdown intensity of silica and sapphire scales as
$I^{cir}/I^{lin}\propto
1/\sqrt[N]{\sigma^{cir}_{N}/\sigma^{lin}_{N}}$, where $\sigma_N$
is the corresponding $N$-photon absorption cross section defining
the rate of the ionization~\cite{Temnov}. The linear polarization
is more efficient in ionization as compared with circular, this
trend is observed up to an irradiance of $\sim
30$~TW/cm$^2$~\cite{Temnov}.

However, once the plasma is formed, the electron quiver energy is
likewise polarization dependent and becomes twice larger for the
\emph{circular} polarization as compared with \emph{linear}
according to $\varepsilon_{osc} \propto (1+\alpha^2)$, where
$\alpha=0, 1$ for linear and circular polarization,
respectively~\cite{Gam,Rode}. The IMP and MPI rates are dependent
on the $\varepsilon_{osc}$ as $w_{mpi}\propto \varepsilon_{osc}^N$
and $w_{imp}\propto \varepsilon_{osc}$; hence, more efficient
ionization is expected for circularly polarized pulses~\cite{Gam}.

The chirp effect at high irradiance is also expected to influence
the ionization rates, especially when photon energy is close to
the fundamental absorption of a target material. By introducing a
negative chirp with shorter wavelength (a larger photon energy) at
the leading part of the pulse, ionization rates can be
significantly enhanced due to $\varepsilon_{osc}^N$ dependence of
$w_{mpi}$. Then, the trailing red-shifted (in wavelength) part of
the pulse will contribute more efficiently to the ionization and
heating of plasma due to the $w_{imp}\propto \lambda^2$
scaling~\cite{Gam,08oe12650}.

Creation of practical X-ray emitting sources depends on
understanding the effects of temporal and spatial
chirp~\cite{Trebino}, polarization, as well as nonlinear phenomena
in delivery of ultra-short laser pulses for X-ray generation at
high irradiance conditions, especially when more complex targets
such as colloidal dispersions and layered materials are used.

\begin{figure*}[tb]
\centering
\includegraphics[width=12.0cm]{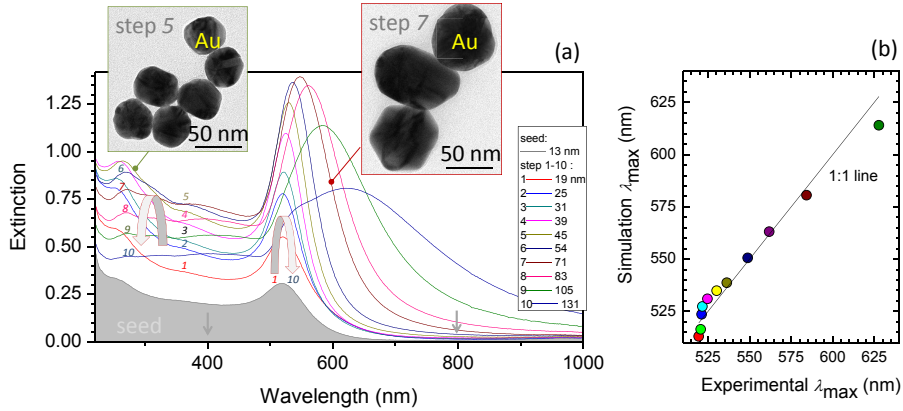}
\caption{(a) Experimental extinction spectra of Au nanoparticle
dispersion at different growth stages (steps from 1 to 10); a 1:3
(sample:water) dilution was used in all spectroscopically
investigated samples as compared to the solution used for X-ray
generation. Insets show TEM images of 45 and 71-nm-diameter
nanoparticles (steps 5 and 7, respectively). The central
wavelength of the fundamental (800 nm) and second harmonics (400
nm) of the fs-laser excitation are marked. The average diameters
of Au nanoparticles were extracted from TEM images for the
different steps of growth. (b) Numerically simulated extinction
spectra of spherical Au nanoparticles; the diameters were
determined experimentally by TEM imaging (see, (a)). Simulations
were carried our using Lumerical FDTD Solutions software. }
\label{f-exp}
\end{figure*}

Here, it is shown that the most efficient X-ray emission  was
observed for Au nanoparticles of $r \simeq 20$~nm radius under
irradiation of negatively pre-chirped pulses which were
approximately 8 times longer compared with the shortest 45~fs
pulses. Chirp can be efficiently used to augment the X-ray
emission by a factor of two over the intensity values observed
using bandwidth limited excitation pulse duration. Power scaling
of X-ray emission implies bulk plasmon generation in gold
nanoparticles and absorption being the driving mechanism of X-ray
emission. Numerical modelling of optical extinction of
nanoparticles separated scattering and absorption contributions
and confirmed the presence of a strong E-field component
perpendicular to the surface, which is required to excite
longitudinal bulk plasmons.

\section*{Results}

\begin{figure}[tb]
\centering
\includegraphics[width=7.50cm]{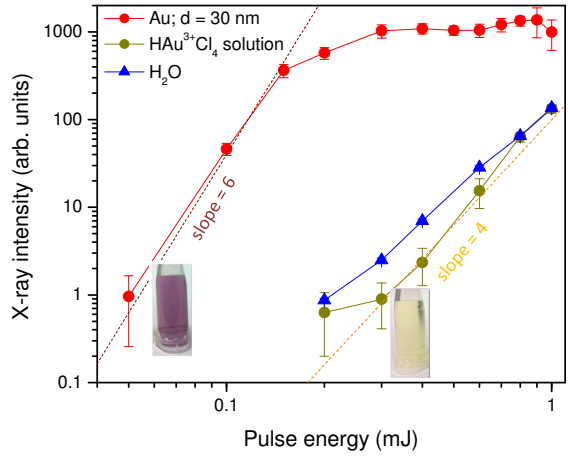}
\caption{Intensity of X-ray emission vs pulse energy for $r =
30$~nm radius Au nanoparticle aqueous dispersion, HAu$^{3+}$Cl$_4$
solution and distilled water (data from different experiments).
Laser pulses of $t_p = 40$~fs duration and $\lambda = 800$~nm
central wavelength were focused with an objective lens of $NA =
0.2$ numerical aperture; insets show images of the solutions. }
\label{f-int}
\end{figure}

\begin{figure}[tb]
\centering
\includegraphics[width=6.0cm]{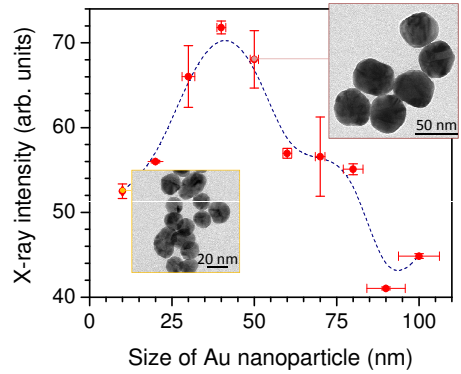}
\caption{X-ray intensity vs Au nanoparticle diameter ($2r$). Au
concentration was 1.2~mmol in all samples. Horizontal error bars
represent the standard deviation of particle diameter and vertical
error bars represent the X-ray intensity measurements performed
three times between 30~min intervals. Pulse energy was $E_p =
0.5$~mJ.} \label{f-xray}
\end{figure}

For the study of X-ray emission and for establishing its power
scaling dependence it was essential to have a very narrow gold
nanoparticle size distribution. For this study we used
mono-dispersed solutions with $< 2\%$ diameter distribution for
the smallest colloidal nanoparticles up to 70~nm in size and $<
6\%$ for the larger ones.

The extinction spectra of dispersions of various size Au
nanoparticles is shown in Fig.~\ref{f-exp} (a). Strong losses at
the UV spectral range do not have a resonant character and are
dominated by the \textit{d}-to-\textit{sp} Au interband
transitions. On the other hand, the plasmonic resonance region
around 520 - 600~nm depends on the nanoparticle size~\cite{Link}.
The strongest relative contribution at the UV wavelengths of
200-300~nm was observed for the 40-50~nm diameter nanoparticles
which proved to facilitate the most efficient generation of
X-rays. The resonant peak positions in extinction spectra were
calculated using experimentally determined nanoparticle sizes for
the step 1-10 generations of the colloidal dispersions.
Figure~\ref{f-exp}(b) shows correlation between the experimental
and numerical peak positions of the plasmonic resonance. In each
case modeling was carried out for a single isolated spherical
nanoparticle. Good correspondence between numerical and
experimental results indicates that nanoparticle solutions had no
agglomerated clusters nor had significant neighboring nanoparticle
interaction effects. This also corroborates that dispersions were
mono-dispersed in terms of nanoparticle size.

\begin{figure}[tb]
\centering
\includegraphics[width=6.50cm]{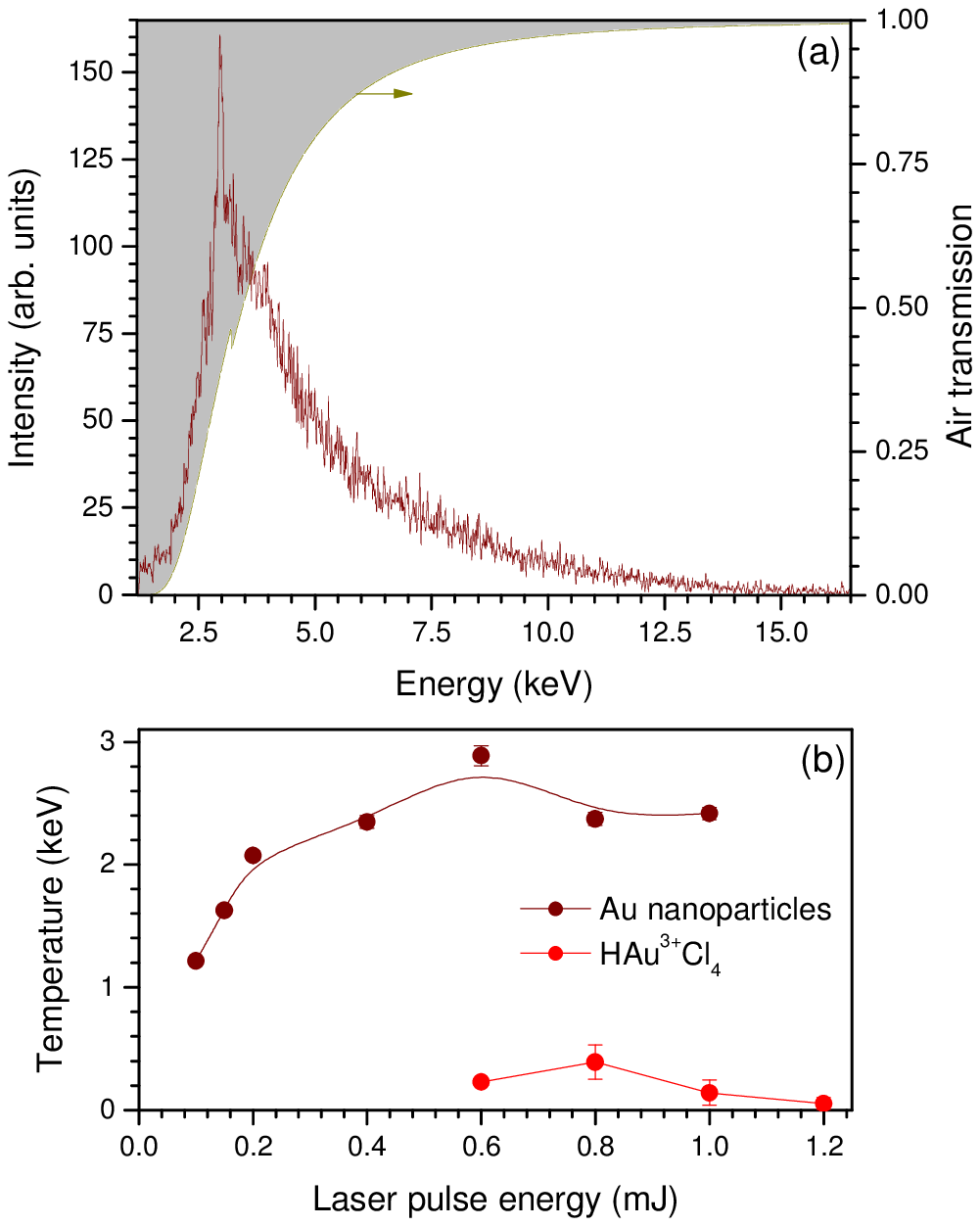}
\caption{(a) X-ray emission spectrum and air transmission. (b)
Temperature of electrons, $T_e$, deduced from the black body fit
of emission from dispersion of Au colloidal particles $2r = (30 -
40)$~nm and 2.5~mM solution of HAu$^{3+}$Cl$_4$.} \label{f-temp}
\end{figure}

Figure~\ref{f-int} shows the power scaling of X-ray generation
from 30-nm-diameter Au nanoparticle dispersion as well as
HAu$^{3+}$Cl$_4$ solution and water for reference. Strong increase
of X-ray intensity is observed for lower pulse energies below
0.1~mJ, with subsequent saturation. The six photon processes can
explain the initial steep power scaling which would correspond to
a 133~nm wavelength. For the electron density in gold $N_e =
5.90\times 10^{22}$~cm$^{-3}$ the bulk plasmon frequency of
$\omega_p = \sqrt{\frac{N_e e^2}{m_e m^*\varepsilon_0}}$
corresponds to the wavelength $\lambda_p^{bulk} = 138$~nm,
considering that the effective electron mass is $m^* = 1$, $m_e$
and $e$ are the free electron mass and charge, respectively.
Usually, the effective electron mass in metals is 20-30\% lower
due to periodicity of the lattice. The 6-photon process, indeed,
is probable for excitation of the bulk plasmon, which is a
longitudinal wave and cannot be directly excited by transverse
E-field of the incident light. However, the bulk plasmon can be
excited by the longitudinal component of the light field which is
comparable with the total field for the 30-50~nm diameter
particles~\cite{11jnn2814}.

\begin{figure}[tb]
\centering
\includegraphics[width=7.50cm]{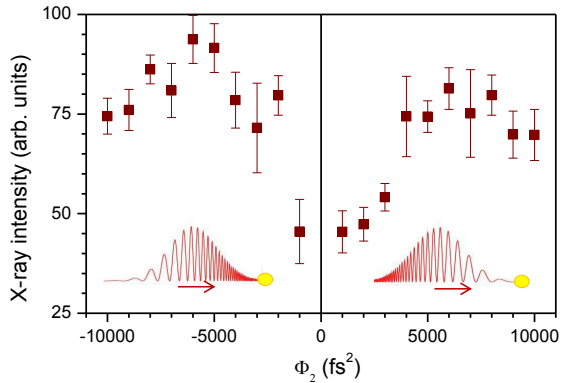}
\caption{Chirp effect: X-ray intensity vs. the second order
dispersion $\Phi_2$~[fs$^2$]; $\Phi_1\equiv 0$. Insets depict
chirped pulse irradiating a gold nanoparticle. Pulse duration
resulting in the strongest X-ray emission at $\Phi_2 =\pm
6000$~fs$^2$ is $t_p = 372$~fs while the shortest pulse was $t_p =
45$~fs at $\Phi_2 = 0$. } \label{f-chirp}
\end{figure}

\begin{figure}[tb]
\centering
\includegraphics[width=8.50cm]{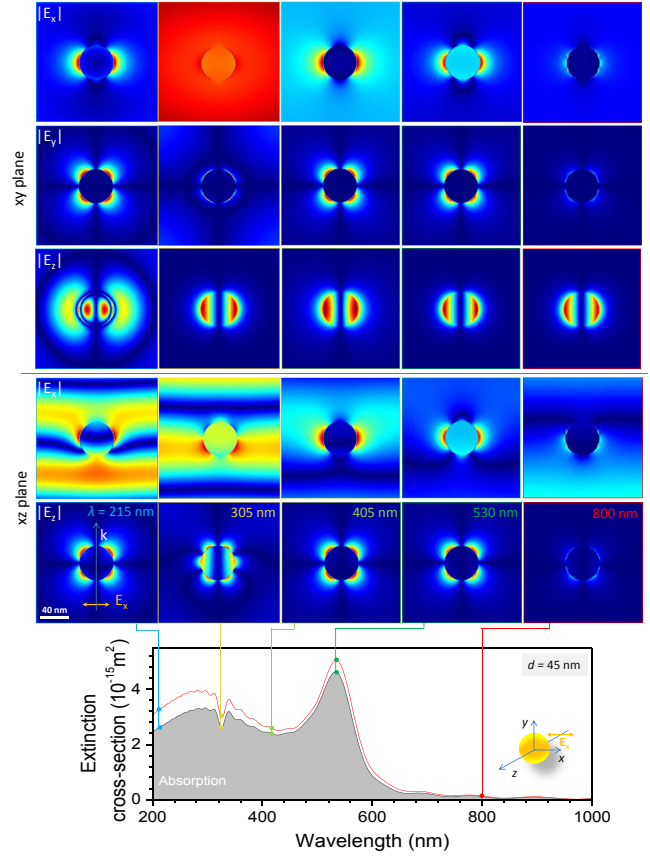}
\caption{The calculated absorption and total extinction spectra
for a $45$~nm diameter Au nano-sphere in water. Schematic E-field
distributions for the different components illustrate the features
of light localisation at several characteristic spectral
locations. The longitudinal $E_z$ field component in xz- and
xy-planes are important for excitation of bulk plasmons; the
incident field is $E_x = 1$, $\mathbf{k}$ is the propagation
direction of light. } \label{f-abs}
\end{figure}

\begin{figure}[tb]
\centering
\includegraphics[width=7.50cm]{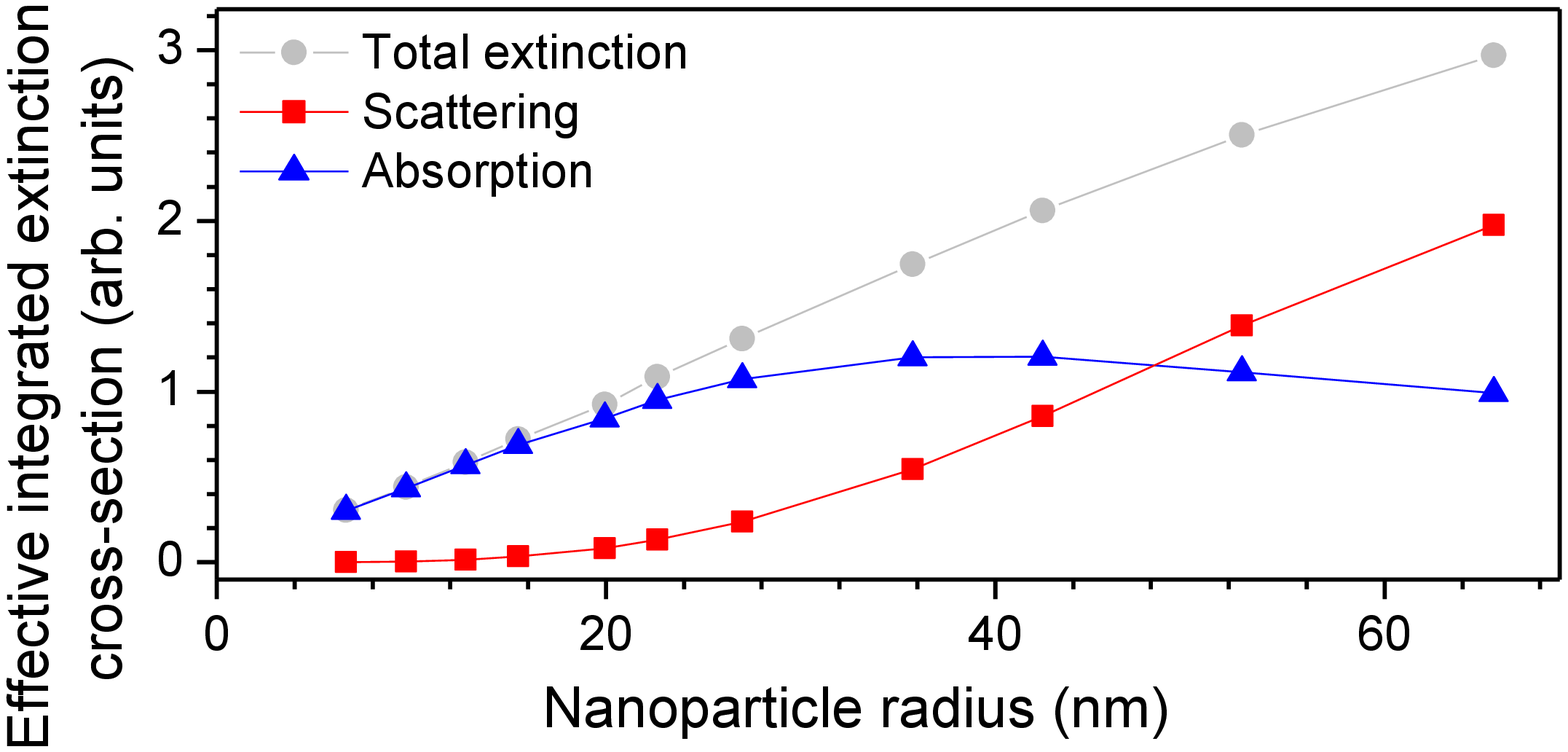}
\caption{The FDTD simulated extinction, absorption, and scattering
cross sections of different size spherical Au nanoparticles,
integrated over the $200$~nm~$\leq \lambda \leq 1000$~nm spectral
span shown in Fig.~\ref{f-abs} and normalized to the corresponding
geometrical cross-sections; $\sigma_{ex} = \sigma_{sc} +
\sigma_{ab}$~[cm$^2$].} \label{f-size}
\end{figure}

The power scaling of X-ray generation in water and
HAu$^{3+}$Cl$_4$ solution (Fig.~\ref{f-int}) exhibits a slope
consistent with a 4-photon process. Interestingly, the 4-photon
absorption would correspond to the $\sim 6.2$~eV energy which was
for a long time assumed as the band gap energy of water and would
be in accordance with water breakdown. However, recently it was
demonstrated that this energy corresponds to the creation of a
solvated electron complex out of the valence band in pure
water~\cite{Linz} while the bandgap of the water is $\sim 9.5$~eV.
The slope of the 4-photon process in water and HAu$^{3+}$Cl$_4$
solution is consistent with breakdown ionisation of water at the
irradiation conditions used.

A systematic study of X-ray emission dependence on particle size
is shown in Fig.~\ref{f-xray}. For the diameters $2r = 40$~nm the
strongest X-ray emission was observed in the saturated region of
the power dependence (Fig.~\ref{f-int}). Intensity of X-ray
emission measured in air is strongly absorbed at the low energy
soft X-ray spectral region (Fig.~\ref{f-temp}(a)). However, when
presence of hard X-rays with photon energy $E > 7$~eV is
substantial, the black body emission fitting allows to extract the
electron temperature $T_e$ from the Planck's formula:
\begin{equation}\label{e2}
I(\nu,T) = \epsilon \frac{2h\nu^3}{c^2}\frac{1}{\exp(h\nu/kT) -
1},
\end{equation}
\noindent where $T$ is the absolute temperature of the emitter,
$h$ is the Planck constant, $\nu = c/\lambda$ is the frequency of
the electromagnetic radiation, $c$ is speed of light in vacuum,
$\lambda$ is the wavelength, and an emissivity $\epsilon \equiv 1$
is assumed for the black body conditions. The results of the
fitting using Eqn.~\ref{e2} are shown in Fig.~\ref{f-temp}(b).
Electron temperatures $T_e \simeq 2.5$~keV  are reached for the Au
nanoparticle dispersion at the most efficient X-ray emission
conditions. An order of magnitude lower emission of X-rays were
observed in a 2.5~mM concentration aqueous solution of
HAu$^{3+}$Cl$_4$.

Influence of the temporal chirp on X-ray emission is presented in
Fig.~\ref{f-chirp}. When $\Phi_2 \simeq \pm 6000$~fs$^2$ the
strongest emission was observed as compared with bandwidth limited
pulse duration of $t_0 = 45$~fs. The $\Phi
 = 6000$~fs$^2$ correspond to a pulse duration $t_p \simeq 372$~fs. Negative chirp is slightly more efficient in X-ray generation as was earlier observed in jets~\cite{08oe12650}.

\section*{Discussion}

Frequencies of the surface plasmon polariton and the bulk plasmon
are related as $\omega_{SP} = \omega_{b}/\sqrt{2}$. For the
$\lambda = 0.8~\mu$m wavelength ($\omega = 2\pi c/\lambda$)
radiation used the 6-photon process, which matches the bulk
plasmon energy in gold, corresponds to $\hbar\omega_b = 9.2$~eV,
whereas surface plasmon polariton energy is $\hbar\omega_{SP} =
6.5$~eV. These plasmon excitations are defined by the electronic
properties of gold and are independent from the particle size and
shape which define the resonances at visible wavelengths (see,
Fig.~\ref{f-exp}). The energy values of 9.2 and 6.5~eV match the
plasma waves in gold and the energy required to create a solvated
electron in water, respectively~\cite{Linz}. The observed power
law of X-ray emission suggests that the light energy absorption
pathway is via the bulk plasmon excitation for gold colloidal
nanoparticles and the breakdown of pure water or solution
(Fig.~\ref{f-int}).

Earlier studies of Au colloidal particles in water under exposure
by 400~nm/150~fs laser pulses showed that nanoparticles with radii
$r = (20-25)$~nm require the smallest fluence to be disintegrated.
At this size nanoparticles absorb enough energy for
dissintegration (explosion) whereas for the smaller nanoparticles
melting occurs. It is noteworthy, that such size does not
correspond to the resonant plasmon band (see, Fig.~\ref{f-exp}).
It was shown recently that explosion of a gold nanoparticle
proceeds via electron ejection from its surface~\cite{Itina}. This
is very similar to the formation of a solvated electron in water
at the initial stage of water breakdown. For the electron
ejection, an E-field component perpendicular to the surface of a
Au nanoparticle is essential. Earlier demonstration of the X-ray
generation via resonant absorption of the longitudinal E-field
component~\cite{08oe12650},
photo-catalysis~\cite{10pss268,11jnn2814,14ct131} and
ablation~\cite{09oe4388} perpendicular to the surface are
consistent with the proposed conjecture that the longitudinal
field component plays a key role in absorption and electron
heating resulting in the X-ray generation from the Au colloidal
particle solution. Even though a strong incident $E_x$ component
is perpendicular to the surface at the equator of a nano-sphere,
it creates a strong reflection and energy is not absorbed, while
the longitudinal component $E_z$ is absorbed~\cite{08bcsj411}
(hence, the resonant absorption).

To reveal the absorption contribution in the total extinction due
to a gold nanosphere in water the cross sections were calculated,
where $\sigma_{ex} = \sigma_{sc} + \sigma_{ab}$
(Fig.~\ref{f-abs}). Furthermore, spatial distributions of E-field
components around particle were extracted. For the linearly
polarised $E_x \equiv 1$ beam incident on the nano-sphere with the
optimum size $2r = 45$~nm for X-ray generation (see,
Fig.~\ref{f-xray}), the total extinction has a contribution of
absorption that is significantly larger than scattering
(Fig.~\ref{f-size}). The spectrum and volume integrated cross
sections, normalized to the geometrical cross-sections of the Au
particles are shown in Fig.~\ref{f-size}. The largest values of an
integral absorption over the range of visible wavelengths mark the
size region of nanoparticles for the most efficient X-ray
generation.

The insets in Fig.~\ref{f-abs} show maps of the E-field
components, including the longitudinal field $E_z$ in the xz- and
xy-planes, for the beam propagating along z-axis
($\mathbf{k}\parallel \mathbf{z}$). The light penetration depth
into the particle is limited to the skin depth, however, there are
peculiar spectral positions for the $E_z$ component for a
particular particle size (see $\lambda = 305$~nm in
Fig.~\ref{f-abs}) where absorption is slightly smaller yet light
field penetration is substantial. Hence, such anti-resonances in
extinction and $E_z$ longitudinal component contribute to the
absorption and energy delivery from the light pulse to a
nanoparticle via resonant absorption (an oscillating dipole
aligned with the linear polarisation of the exciting beam would
increase reflection and reduce absorption~\cite{08bcsj411}). Due
to its longitudinal character, this $E_z$ field can excite a bulk
plasmon. The $E_z$ component in the xy-plane reveals its strong
presence inside the nanoparticle which is expected to cause
substantial absorption.

Generation of hard X-rays from a 2.5~keV temperature plasma
described here is an efficient process in this part of EM
spectrum. Conversely, for 1~THz (300~$\mu$m) radiation the
Eqn.~\ref{e2} predicts a $4.6\times 10^{11}$ times smaller
radiance as calculated for the black body heated to 5~keV (with
peak emission at 2.46$\mathrm{\AA}$). The emitted photon number
ratio, between X-rays and T-rays at correspondingly
2.46$\mathrm{\AA}$ and 300~$\mu$m wavelength is, however,
significantly smaller $3.8\times 10^{5}$. Hence, terahertz
emission from breakdown plasmas can be intense enough for use in
applications~\cite{tray,tray1}, however, in the case of water
solutions a back-scattering geometry has to be used for the T-ray
source due to strong water absorption.

\section*{Conclusions}

Hard X-ray emission from aqueous solution jets of 30-50~nm
diameter Au nanoparticles is demonstrated to follow a 6-photon
process consistent with bulk plasmon excitation in gold. Formation
of the longitudinal component of the incident light field $E_z$
($z$ is the propagation direction $\mathbf{k} = 2\pi/\lambda$)
required for excitation of bulk plasmon is due to near-field
focusing by gold colloidal nanoparticles. Electron temperatures
corresponding to the black body radiation of 2.5~keV produce a
bright hard X-ray source which can be used in air conditions.
Temporal chirp of 45~fs pulses provides a tool to control X-ray
intensity. This mechanism is additionally corroborated by the
X-ray generation scaling power law for water and is consistent
with electron ejection from water molecule and formation of
solvated electron at $\sim 6.5$~eV~\cite{Linz}.

Creation of high temperature plasma fragmentation of colloidal
particles in solution opens rich possibilities to investigate
temporal dynamics of the breakdown on sub-wavelength scales and at
short time windows where modeling is already predicting different
scenarios to be cross checked and validated
experimentally~\cite{Itina,Linz}. Laser-driven explosions with
high temperature black body radiation can provide not only hard
X-ray sources but also efficient THz emitters. Further studies are
strongly required for directionality control for such EM-radiation
to be used in practical applications.

\section*{Methods}

Ultra-short laser pulses were used for X-ray generation: pulse
duration was $t_0 = 45$~fs at the central wavelength of $\lambda =
800$~nm, pulse energy typically ranged from $E_p = 0.1$~mJ to
1~mJ, at 1~kHz repetition rate. Focusing was carried out with an
off-axis parabolic mirror with focal length $f = 5$~cm in air at
atmospheric pressure; the equivalent numerical aperture of
focusing was $NA = 0.2$. Pulse can be described by a Gaussian
temporal envelope of the form $E(t) = E_0e^{-2\ln 2
(t/t_p)^2}\cos(\omega t + \beta t^2)$, where $\omega$ is the
cyclic frequency of light, $t_p$ is the pulse duration at the
full-width at half maximum (FWHM), $\beta$ [1/fs$^2$] is the
linear chirp, and $E_0 = \sqrt{2I_0/(c\varepsilon_0n)}$ is the
field amplitude, $n$ is the refractive index, $c$ is speed of
light, $t$ is time, $I_0 = 2I_{av}$ is the peak intensity which is
twice larger than the average, $I_{av}$ for the Gaussian, and
$\varepsilon_0$ is the permittivity of vacuum. The instantaneous
cyclic frequency $\omega_{ins}(t) =\omega_0 +2\beta t$, where
$\beta>0$ corresponds to the positive chirp with trailing high
frequency components.

Pulse pre-chirping is implemented to increase pulse duration up to
ten times using a 1D array of liquid crystal cells (FemtoJock,
Biophotonics Solutions, Inc.). For the slowly with frequency
$\omega$ varying spectral phase $\varphi(\omega)$ it can be
expanded into the Taylor series around the central frequency
$\omega_0$ with the few first terms as $\varphi(\omega) =
\varphi(\omega_0) + \Phi_1\omega + \frac{\Phi_2}{2!}\omega^2+...$,
where $\varphi(\omega_0)$ is the absolute phase of the pulse in
time domain, the first derivative $\varphi'(\omega_0)=\Phi_1$ is
the group delay (GD) which defines a shift of envelope in time
domain, $\varphi''(\omega_0)=\Phi_2$~[fs$^2$] is the group delay
dispersion (GDD) or the second order dispersion which defines the
chirp in time domain. Duration of the time broadened pulse at FWHM
is defined as $t_p = t_0\sqrt{1+\left[4\ln2\Phi_2/t_0^2\right]^2}
\equiv t_0\sqrt{1+\beta^2}$ where $t_0 = 45$~fs is the shortest
spectral bandwidth limited pulse duration. For this study higher
orders as well as $\Phi_1$ were set to zero.

The pulse energy of $E_p = 0.1$~mJ corresponds to the field
strength $E_0 = 4.23\times 10^{11}$~V/m, which is much higher than
the breakdown of dry air at $6\times 10^6$~V/m. This complicates
the determination of the actual fluence and irradiance on the
nanoparticle due to breakdown of air and water jet with colloidal
particles as discussed in detail in the Sec.~\blue{Discussion}.
The power scaling of X-ray emission is used instead to reveal the
mechanism of X-ray generation rather the incident
fluence/irradiance onto a nanoparticle.

Gold nanoparticle colloidal aqueous dispersions with the Au atomic
concentrations at $1.2\times 10^{-3}$~mol/L with different
particle diameters ranging from 10 to 100~nm were used for X-ray
generation. Nanoparticles were negatively charged and surfactant
coated to prevent agglomeration. Figure~\ref{f-exp} shows optical
extinction, cumulative scattering and absorbtion losses, spectra
measured in transmission. A water jet 1~mm in diameter was formed
at the flow rate of 2.5~m/s and a laser beam was delivered onto to
the front surface of the jet. Optimisation of the focal region
placement was carried out for the highest yield of X-ray emission.

X-ray emission intensity measurements were carried out using a
Geiger counter (model5000, Health Physics Instrument, Inc.). X-ray
emission spectroscopy was performed using a solid-state detector
(XR-100CR, PX2CR, Amptek).

Numerical modeling of extinction spectra and light field
enhancement were carried out using finite difference time domain
(FDTD) software package (FDTD Solutions, Lumerical Solutions
Inc.). A uniform $0.5$~nm 3D mesh was used, permittivity of gold
was tabulated in software according to results reported by Johnson
and Christy~\cite{jcau}.


\section*{Author contributions statement}
This study was primarily designed and conducted by K.H., F.C.P.M.,
and W.H.H, as the principal investigators, provided conceptual and
technical guidance in all aspects of the research project. K.H.
and F.C.P.M. planned and performed the X-ray intensity and X-ray
emission spectra measurements. M.P., M.T.N., and T.Y. contributed
the preparation and characterization of gold nano-colloidal
solutions. A.B. and S.J. performed and analyzed the calculated
absorption/extinction measurements and FDTD simulations. The
manuscript was written by K.H. and commented on by all authors.

\section*{Additional Information}
The authors declare no competing financial interests.

\section*{Acknowledgements}
S.J. is grateful for partial support via the Australian Research
Council DP130101205 Discovery project. FDTD simulation work was
performed on the swinSTAR supercomputer at Swinburne University of
Technology.

%

%
%

\end{document}